\documentclass{article}
\usepackage{graphicx}
\usepackage{caption}
\usepackage{subcaption}

\usepackage{geometry}
\usepackage[utf8]{inputenc}
\usepackage{amsmath,mathrsfs,amssymb}
\usepackage{physics}
\usepackage{tensor}
\usepackage[inline]{enumitem}
\usepackage{xcolor} 

\newcommand{\kdelta}[1]{\delta\indices{#1}}

\newcommand{\strconst}[1]{C\indices{#1}}
\newcommand{\connection}[1]{\Gamma\indices{#1}}

\newcommand{\curvature}[1]{R\indices{#1}}

\usepackage{url}
\usepackage{authblk}
\usepackage{biblatex} 
\usepackage{graphicx} 

\usepackage{geometry}
\usepackage[utf8]{inputenc}
\usepackage{amsmath,mathrsfs,amssymb}
\usepackage{physics}
\usepackage{tensor}
\usepackage[inline]{enumitem}

\DeclareMathOperator{\diag}{diag}
\usepackage{xcolor}

\title{Cosmic Acceleration from topological considerations III: Lie group
}

\author[1]{I. A. Sarmiento-Alvarado\thanks{ignacio.sarmiento@cinvestav.mx}}
\author[2]{Maribel Hernández-Márquez\thanks{maribel.hernandez@nucleares.unam.mx}}
\author[1]{Tonatiuh Matos\thanks{tonatiuh.matos@cinvestav.mx}}
\affil[1]{Departamento de F\'{\i}sica, Centro de Investigaci\'on y de Estudios Avanzados del IPN, Av. I.P.N. 2508, San Pedro Zacatenco, M\'exico 07360, CDMX.}
\affil[2]{Instituto de Ciencias Nucleares, Universidad Nacional Autónoma de México, Apdo. Postal 70-543, Deleg. Coyoacán, C.P. 04510, CDMX, México.}

\date{}

\begin{document}
\maketitle
\begin{abstract}
Recent observations on the large-scale structure of the universe indicate that the cosmological constant cannot be the definitive answer to the nature of dark energy. Therefore, it is a good time to propose alternatives to understand the late-accelerated expansion of the universe.
      In this work we study the possibility that the acceleration of space-time is due to the topology of the universe.
      We assume that the topology of the universe is a principal fiber bundle whose base space is our 4-dimensional spacetime and the fiber is an $N$-dimensional Lie group that evolves with time. For the base space we consider a homogeneous and isotropic spacetime, we find that the base space is currently accelerating for $1 < N$ compact semi-simple Lie groups whose scale-factors are equal and  for $1 < N$ non-Abelian Lie groups whose scale-factors are different and as long as its structure constants satisfy some conditions.\\
      However when we study the evolution of the density parameters these differ from the evolution within the  $\Lambda CDM$ model, this led us to think in the possibility of use a different group as fiber in order to obtain the right evolution of the density parameters. We conclude that it is possible that the accelerated expansion of the universe is due to consider a different topology of the universe as a principal fiber bundle.\\
\end{abstract}

\section{Introduction}
Since the discovery of the late-time accelerating expansion of the universe in 1998 by means of Supernovae observations \cite{perlmutter1999measurements},\cite{riess1998observational}  there have been different proposals trying to explain it. Within the theoretical framework of general relativity the most accepted explanation for this acceleration is the existence of a cosmological constant that can be described as a perfect fluid whose equation of state is $\omega_{\Lambda}=-1$, although this can explain the late-time acceleration of the universe, the nature and the small value of the cosmological constant is still a mystery \cite{martin2012everything}. Furthermore the discrepancy of the values obtained for the the Hubble constant from the CMB and local observations \cite{riess2022comprehensive} have led us to think that the equation of state of dark energy is dynamical \cite{escamilla2022dynamical},\cite{zhao2017dynamical}, and more recently from the results of the observations carried out by DESI \cite{DESI:2024wki} we can conclude that it is difficult for the cosmological constant to be the last answer to the nature of dark energy.

In order to explain the late-accelerated expansion of the universe, in a previous work \cite{Hernandez-Marquez_2020} it was proposed that the topology of the whole universe is a principal fiber bundle where  the  base space is the four-dimensional space-time and the fiber is an N-dimensional Lie group. In the same way that the base space is expanding and this is described mathematically by means of the evolution of the scale factor $a(t)$ in \cite{Hernandez-Marquez_2020} we considered that the fiber also evolves with time and this evolution is described by their corresponding scale factors of the fiber. With this hypothesis and assuming that the $N+4$-dimensional Einstein field equations are valid it was found that if the fiber is the 3-dimensional Lie group $SU(2)$ then there are initial conditions that allows the universe starts to accelerate at some redshift $z$. Furthermore in \cite{Hernandez-Marquez_2020} it was defined the equation of state of dark energy $\omega_e$ and it was found that the current value of $\omega_e$, considering the group $SU(2)$ as fiber is less than $-1$; also this equation of state is dynamical.\\

In this work we follow the hypothesis proposed in \cite{Hernandez-Marquez_2020}, but now, we consider the topology of the universe as a principal fiber bundle where the base space is our 4-dimensional space-time and we study the general case in which the fiber is an $N$-dimensional Lie group, with $N\geq 1$ and not only the specific case in which the fiber is the group SU(2). Therefore we investigate what happens with the expansion of the universe when considering fibers other than the $SU(2)$ group and of larger dimensions.  For that purpose in section \ref{section:fieldequations} we obtain the field equations of our model in terms of the redshift. In section \ref{section:energy-momentum} we obtain the evolution of density parameters of matter and radiation. In section \ref{section:omegae} we find the parameter density of what we call in our model "dark energy". Then in the following sections we find the evolution of the Hubble parameter, deceleration parameter and the scale factors of the fibers, for specific cases. In section \ref{onedimensional}, we study the case where the fiber is a one-dimensional Lie group. 
For the $N$-dimensional Lie group, in section \ref{oneparameter}, we consider a left-invariant metric where all the scale factors $b_i(t)$ are equal  and also in section \ref{section9:differentscale} we study the case when all scale factors are different. In section \ref{section:evolutiondensity}, we describe mathematically the evolution of the density parameters. 
Finally in section \ref{conclusions} we discuss our results and we give our conclusions. 

Throughout this paper, the indices $A, B, C, D$ take the values from 1 to $N + 4$ and the indices $i, j, k, l, m$ from 5 to $N + 4$.
$\dot f$ and $\ddot f$ denote the first and second derivative of $f$ with respect to time $t$, respectively.

\section{Geometry}  
\label{section:geometry}
We consider that our whole universe is composed by a 4-dimensional base space and a fiber forming a principal fiber bundle.
The spacetime is represented by the 4-dimensional base space and we assume that is homogeneous and isotropic.
Thus for the base space we consider the Friedman-Lemaître-Robertson-Walker metric, that is,
\begin{align}
    g_{FLRW} = \dd t \otimes \dd t
    - \frac{ a ^2 (t) }{ 1 - K r ^2 } \dd r \otimes \dd r
    - a ^2 (t) r ^2 \dd \theta \otimes \dd \theta
    - a ^2 (t) r ^2 \sin ^2 \theta \dd \phi \otimes \dd \phi
\end{align}
where $a$ is the scale factor and $K$ is a constant representing the spatial curvature of the spacetime.\\
For the fiber we use a left-invariant metric and 
according to \cite{Hernandez-Marquez_2020} if we ignore the interactions between particles this can be written as 
\begin{align}
    g _{ \textit{Lie} } = \sum _i b _i ^2 (t) \omega ^i \otimes \omega ^i
\end{align}
where $\{ \omega_i \}$ is a basis for the left-invariant 1-forms of the group and $b_i$ are functions which depend on $t$. In analogy with $a(t)$, the functions $b_i(t)$ are the scale factors of the fiber and therefore these spatial extra-dimensions are evolving with time. \\
Therefore, the metric for the whole universe it can  be written as
\begin{align}
\label{metric tensor}
    \bar g = \dd t \otimes \dd t - \frac{ a ^2 (t) }{ 1 - K r ^2 } \dd r \otimes \dd r - a ^2 (t) r ^2 \dd \theta \otimes \dd \theta - a ^2 (t) r ^2 \sin ^2 \theta \dd \phi \otimes \dd \phi - \sum _i b _i ^2 (t) \omega ^i \otimes \omega ^i.
\end{align}
And in terms of a rigid basis the metric tensor \eqref{metric tensor} can be written as
\begin{equation}
    \bar g = \eta_{A B} e ^A \otimes e ^B
\end{equation}
where
\begin{align}
    \eta_{A B} = \diag( 1 , -1 , -1 , -1 , \underbrace{ -1 , \ldots , -1 } _\text{$N$ times} ) \ ,
\end{align}
\begin{equation}
    \left\{ e^A \right\} = \left\{ e ^1 = \dd t, \, e ^2 = \frac{ a }{\sqrt{ 1 - K r ^2 }} \dd r, \, e ^3 = a r \dd \theta, \, e ^4 = a r \sin \theta \dd \phi, \, e ^i = b _i \omega ^i \right\},
\end{equation}
is a dual basis of an orthonormal basis $\{ e_A \}$.\\

To obtain the components of the Ricci tensor we need to compute the exterior derivative of each 1-form of the dual basis $\{ e^A \}$.
In the case of the 1-forms $\omega ^i$ of a Lie group they satisfy
\begin{equation}
\label{domega}
    \dd \omega ^i = - \frac{1}{2} \sum _{j,k} \strconst{^i _j _k} \omega ^j \wedge \omega ^k,
\end{equation}
where $ \strconst{^i _j _k} $ are structure constants with respect to the basis for the left-invariant vectors $\{X_i\}$. And according to \cite{10.21468/SciPostPhysLectNotes.21}
\begin{align}
\label{antisymmetry structure constants}&
    \strconst{^i _j _k} = - \strconst{^i _k _j},
\\\label{Jacobi identity}&
        \sum _m \qty(
        \strconst{^m _i _j} \strconst{^l _m _k}
        + \strconst{^m _j _k} \strconst{^l _m _i}
        + \strconst{^m _k _i} \strconst{^l _m _j}
    ) = 0.
\end{align}
Then
\begin{equation}
\begin{aligned}
    \dd e ^2 & = \frac{ \dot a }{ a } e ^1 \wedge e ^2, \\
    \dd e ^3 & = \frac{ \dot a }{ a } e ^1 \wedge e ^3 + \frac{\sqrt{ 1 - K r ^2 }}{ a r } e ^2 \wedge e ^3, \\
    \dd e ^4 & = \frac{ \dot a }{ a } e ^1 \wedge e ^4 + \frac{\sqrt{ 1 - K r ^2 }}{ a r } e ^2 \wedge e ^4 + \frac{ \cot \theta }{ a r } e ^3 \wedge e ^4, \\
    \dd e ^i & = \frac{ \dot b _i }{ b _i } e ^1 \wedge e ^i -  \frac{1}{2} \sum _{j,k} \strconst{^i _j _k} \frac{ b _i }{ b _j b _k } e ^j \wedge e ^k,
\end{aligned}    
\end{equation}
where we have used Eq. (\ref{domega}) in the last expression to find $\dd e^i$.\\ 
From the Cartan first structural equations
\begin{align}
    \dd e ^A = \connection{^A _[ _B _C _] } e ^B \wedge e ^C,
\end{align}
we get the Ricci rotation coefficients
\begin{equation}
\begin{aligned}
    \connection{^2 _[ _1 _2 _]} & = \connection{^3 _[ _1 _3 _]} = \connection{^4 _[ _1 _4 _]} = \frac{ \dot a }{ 2 a },
\\
    \connection{^3 _[ _2 _3 _]} & = \connection{^4 _[ _2 _4 _]} = \frac{\sqrt{ 1 - K r ^2 }}{ 2 a r },
\\
    \connection{^4 _[ _3 _4 _]} & = \frac{ \cot \theta }{ 2 a r },
\\
    \connection{^i _[ _1 _i _]} & = \frac{ \dot b _i }{ 2 b _i },
\\
    \connection{^i _[ _j _k _]} & = - \frac{\strconst{^i _j _k}}{2}\frac{b_i}{b_jb_k}.
\end{aligned}    
\end{equation}
The connection 1-form is defined as
\begin{equation}
    \connection{^A _B} = \connection{^A _B _C} e ^C
\end{equation}
Using the identity
\begin{equation}
        \connection{_A _B _C} = \connection{_A _[ _B _C _]} + \connection{_B _[ _C _A _]} - \connection{_C _[ _A _B _]},
\end{equation}
we find
\begin{equation}
\label{connection one-form}
\begin{aligned}
    \connection{^1 _2} & = \frac{ \dot a }{ a } e ^2,    &
    \connection{^1 _3} & = \frac{ \dot a }{ a } e ^3,    &
    \connection{^1 _4} & = \frac{ \dot a }{ a } e ^4,    \\
    \connection{^2 _3} & = -\frac{\sqrt{ 1 - K r ^2 }}{ a r } e ^3,  &
    \connection{^2 _4} & = -\frac{\sqrt{ 1 - K r ^2 }}{ a r } e ^4,  &&&\\
    \connection{^3 _4} & = -\frac{ \cot \theta }{ a r } e ^4,    &&&&&\\
    \connection{^1 _i} & = \frac{ \dot b _i }{ b _i } e ^i,  &
    \connection{^i _j} & = - \frac{1}{2} \sum _k \strconst{^i _j _k} \frac{ b _i ^2 + b _j ^2 - b _k ^2}{ b _i b _j b _k } e ^k. 
\end{aligned}    
\end{equation}
The Cartan second structural equations are
\begin{equation}
    \mathcal{R}\indices{^A _B} 
    = \dd \connection{^A _B} + \connection{^A _C} \wedge \connection{^C _B}
    = \frac{1}{2} \curvature{^A _B _C _D} e ^C \wedge e ^D.
\end{equation}
For our connection 1-form \eqref{connection one-form} we have that the non-zero components of Riemann tensor are
\begin{equation}
\begin{aligned}
    \curvature{^1 _2 _1 _2} & = \curvature{^1 _3 _1 _3} = \curvature{^1 _4 _1 _4} = \frac{ \ddot a }{ a },
\\
    \curvature{^2 _3 _2 _3} & = \curvature{^2 _4 _2 _4} = \curvature{^3 _4 _3 _4} = \frac{ \dot a ^2 + K }{ a ^2 },
\\
    \curvature{^1 _i _1 _j} & = \frac{ \ddot b _i }{ b _i} \delta_{i j},
\\
    \curvature{^2 _i _2 _j} & = \curvature{^3 _i _3 _j} = \curvature{^4 _i _4 _j} = \frac{ \dot a \dot b _i }{ a b _i } \delta_{i j},
\\
    \curvature{^1 _i _j _k} & = \frac{\strconst{^i _j _ k}}{ 2 b _i b _j b _k } \qty[
        b _i ^2 \qty(
            \frac{ \dot b _j }{ b _j }
            + \frac{ \dot b _k }{ b _k }
            - 2 \frac{ \dot b _i }{ b _i }
        )
        + \qty( b _j ^2 - b _k ^2 ) \qty(
            \frac{ \dot b _j }{ b _j }
            - \frac{ \dot b _k }{ b _k }
        )
    ],
\\
    \curvature{^i _j _k _l} & = \frac{ \dot b _i \dot b _j }{ b _i b _j } \qty( \kdelta{^i _k} \kdelta{^j _l} - \kdelta{^i _l} \kdelta{^j _k} )
    + \sum _m \strconst{^i _m _j} \strconst{^m _k _l} \frac{b_m^2 - b_i^2 - b_j^2}{2 b _i b _j b _k b _l}
\\& + \sum _m \strconst{^i _m _k} \strconst{^m _j _l} \frac{( b_m^2 + b_i^2 - b_k^2 ) ( b_m^2 + b_j^2 - b_l^2 )}{4 b _i b _j b _k b _l b _m ^2}
\\& - \sum _m \strconst{^i _m _l} \strconst{^m _j _k} \frac{( b_m^2 + b_i^2 - b_l^2 ) ( b_m^2 + b_j^2 - b_k^2 )}{4 b _i b _j b _k b _l b _m ^2}.      
\end{aligned}
\end{equation}
where $\kdelta{^i _j}$ is the Kronecker delta.
The Ricci tensor is the contraction of the curvature tensor
\begin{equation}
    \curvature{_A _B} = \curvature{^C _A _C _B}.
\end{equation}
Its non-zero components are
\begin{align}
\label{R11}
    \curvature{_1 _1}
&   = -3 \frac{ \ddot a }{ a } - \sum _i \frac{ \ddot b _i }{ b _i }
\\\label{R22}
    \curvature{_2 _2}
&   = \curvature{_3 _3} = \curvature{_4 _4} = \frac{ \ddot a }{ a } + 2 \frac{ \dot a ^2 + K }{ a ^2 } + \frac{ \dot a }{ a }\sum _i \frac{ \dot b _i }{ b _i }
\\\label{R1j}
    \curvature{_1 _j}
&   = - \sum _k \strconst{^k _j _k} \frac{b_j}{2 b_k^2} \left( \frac{\dot b_j}{b_j} - \frac{\dot b_k}{b_k}, \right)
\\\label{Rij}
\curvature{_i _j}
&   = \qty(
            \frac{ \ddot b _i }{ b _i }
            + 3 \frac{ \dot a \dot b _i }{ a b _i }
            + \frac{ \dot b _i }{ b _i } \sum _k \frac{ \dot b _k }{ b _k }
            - \frac{ \dot b _i ^2 }{ b _i ^2 }
        ) \delta_{i j}
    + \sum_{k, l} \strconst{^k _l _k} \strconst{^l _i _j} \frac{b_l^2 + b_i^2 - b_j^2}{4 b_i b_j b_k^2}
\\& + \sum_{k, l} \strconst{^k _l _i} \strconst{^l _k _j} \frac{(b_k^2 - b_l^2 + b_i^2) (b_k^2 - b_l^2 - b_j^2)}{4 b_i b_j b_k^2 b_l^2}.
\end{align}
The Ricci scalar is the trace of the Ricci tensor
\begin{equation}
\label{scalar curvature}
    \curvature{} = g^{A B} \curvature{_A _B}.
\end{equation}
Replacing the Eqs. \eqref{R11}, \eqref{R22},\eqref{Rij} in Eq.(\ref{scalar curvature}), we get:
\begin{equation}
\begin{aligned}
    \curvature{} & = -6 \frac{ \ddot a }{ a }
        - 2 \sum _i \frac{ \ddot b _i }{ b _i }
        - 6 \frac{ \dot a ^2 + K }{ a ^2 }
        - 6 \frac{ \dot a }{ a } \sum _i \frac{ \dot b _i }{ b _i }
        + \sum _i \frac{ \dot b _i ^2 }{ b _i ^2 }
        - \qty( \sum _i \frac{ \dot b _i }{ b _i } ) ^2
\\&     + \sum_i \frac{B_{ii}}{2 b_i^2}
        - \sum _{i,j,k} \strconst{^j _k _i} \strconst{^k _j _i} \frac{ b _j ^4 + b _k ^4 - b _i ^4 }{4 b _i ^2 b _j ^2 b _k ^2 },
\end{aligned}
\end{equation}
where
\begin{equation}
    B_{i j} = \sum_{k,l} \strconst{^k _l _i} \strconst{^l _k _j}.
\end{equation}
is the Cartan-Killing metric.

\section{Energy-momentum tensor} 
\label{section:energy-momentum}  
In this work we consider an energy-momentum tensor given by
\begin{equation}
\label{energy momentum tensor}
    T_{A B} = \rho e^1 \otimes e^1 + p ( e^2 \otimes e^2 + e^3 \otimes e^3 + e^4 \otimes e^4 ),
\end{equation}
and from the conservation of the energy momentum tensor
\begin{equation}
\label{mass eq}
    \dot \rho
    + \qty( 3 \frac{ \dot a }{ a }
        + \sum _i \frac{ \dot b _i }{ b _i }
    ) \rho
    + 3 \frac{ \dot a }{ a } p
    = 0,
\end{equation}
and assuming
\begin{equation}
    p = \omega \rho,
\end{equation}
with $\omega$ as a real constant, we find that in terms of the redshift $z$
\begin{equation}
    \rho = \rho _o (1 + z) ^{ 3 ( \omega + 1 ) } V
\end{equation}
where $\rho_0 = \rho (z = 0)$, 
\begin{equation}
\label{V}
    V = \prod _i \frac{ b _{i 0} }{ b _i }
\end{equation}
and $b _{i 0} = b_i (z = 0)$.
By means of the critical density
\begin{equation}
\label{critical density}
    \rho _c = \frac{ 3 H ^2 }{ 8 \pi G },
\end{equation}
we introduce the density parameter
\begin{equation}
    \Omega = \frac{ \rho }{ \rho _c },
\end{equation}
then
\begin{equation}
\label{def density parameter}
    \Omega = \Omega_0 \qty( 1 + z ) ^{ 3 ( \omega + 1 ) } \frac{H_0 ^2}{H ^2}V,
\end{equation}
where $8\pi G=\kappa_4$ is the four-dimensional gravitational constant,
\begin{equation}
\label{def H}
    H = \frac{\dot a}{a}
\end{equation}
is the Hubble parameter, $H_0 = H (z = 0)$ and $\Omega_o = \Omega (z = 0)$.\\
We consider a total density
\begin{equation}
\label{total density}
    \rho = \rho _m + \rho _\gamma
\end{equation}
where $\rho_m$ is the density of the baryonic matter plus dark matter and $\rho_\gamma$ is the density of radiation.
Its corresponding density parameters are
\begin{align}
\label{desinty parameter matter}
    \Omega_m & = \Omega_{m 0} \qty( 1 + z ) ^3 \frac{H_0 ^2}{H ^2} V
\\\label{desinty parameter radiation}
    \Omega_\gamma & = \Omega_{\gamma 0} \qty( 1 + z ) ^4 \frac{H_0 ^2}{H ^2} V
\end{align}
being $\Omega_{m 0} = \Omega_m ( z = 0 )$ and $\Omega_{\gamma 0} = \Omega_\gamma ( z = 0 )$.

\section{Field equations}       
\label{section:fieldequations}  
We assume that the Einstein field equations are satisfied in a space-time of ${N}+4$ dimensions, then  
\begin{equation}
\label{EFE}
    \curvature{_A _B} - \frac{ \curvature{} }{2} \eta_{A B} = \kappa_h T_{A B},
\end{equation}
where $k_h$ is the gravitational constant for the space-time of ${N}+4$ dimensions, the above equation can be rewritten as:
\begin{align}
\label{field eqs}
    \curvature{_A _B} = \kappa_h \qty( T_{A B} - \frac{T}{ N + 2 } \eta_{A B} ),
\end{align}
where $T$ is the trace of the energy momentum tensor and ${N}$ is the dimension of the Lie group. It is important to mention that in \cite{Hernandez-Marquez_2020}  it was considered $\kappa_h=\kappa_4$, where $\kappa_4$ is the four-dimensional gravitational constant but in this work we consider that it could be different and we define the ratio $ \kappa=\kappa_h/\kappa_4.$  \\
It is convenient to write the field equations in terms of redshift parameter $z$ instead the time $t$.
To do it, we use the identities
\begin{align}
\label{1st derivative}
    \dot f & = - (1 + z) H f ^\prime  \\
\label{2do der}
    \ddot f & = (1 + z)^2 H^2 \left( f ^{\prime\prime} + \left[ \frac{1}{1 + z} + \frac{H^\prime}{H}\right] f^\prime \right)
\end{align}
Note that $f^\prime$ and $f^{\prime\prime}$ denote the first and second derivatives of the function $f$ with respect to $z$, respectively.
Differentiating \eqref{def H} with respect to time $t$ and using Eq. \eqref{1st derivative} we obtain
\begin{equation}
\label{prime H}
    \frac{H^\prime}{H} = \frac{1 + q}{1 + z}
\end{equation}
where
\begin{equation}
\label{def q}
    q = - \frac{ \ddot a }{ a H ^2 }
\end{equation}
is the deceleration parameter \cite{2015arXiv150200811B}.
Substituting the Eq. \eqref{prime H} in Eq. \eqref{2do der} we rewrite $\ddot f$ as
\begin{equation}
\label{2do derivative}
    \ddot f = (1 + z)^2 H^2 \left( f ^{\prime\prime} + \frac{2 + q}{1 + z} f^\prime \right).
\end{equation}
Then from $R_{11}=\kappa_{h}T_{11}$, we obtain 
\begin{equation}
\label{field eq R11}
    3 q - (1 + z)^2 \sum_i I_i^\prime 
    = 3 \kappa \left( \frac{N + 1}{N + 2} \Omega_m + \Omega_\gamma \right) + (1 + z) (2 + q) \sum_i I_i + (1 + z)^2 \sum_i I_i^2,
\end{equation}
from $R_{22}=\kappa_hT_{22}$
\begin{equation}
\\\label{q}
    q
  = 2 - \kappa \left( \frac{3 \Omega_m}{N + 2} + \Omega_\gamma \right) - 2 \Omega_K - (1 + z) \sum_i I_i
\end{equation}
and finally from $R_{ii}=\kappa_hT_{ii}$, we get
\begin{equation}
\label{Ii'}
    I_i^\prime = \left[
        \kappa \left( \frac{3 \Omega_m}{N + 2} + \Omega_\gamma \right) + 2 \Omega_K - 1
    \right] \frac{I_i}{1 + z}
    + \left[
        \frac{3 \kappa \Omega_m}{N + 2} - \frac{u _i}{H^2}
    \right] \frac{1}{(1 + z)^2},
\end{equation}
where
\begin{align}
\label{def Ii}
    I_i & = \frac{b_i^\prime}{b_i},
\\
    u_i & = -\frac{B_{ii}}{2 b_i ^2} + \sum _{j,k} \strconst{^k _l _i} \strconst{^l _k _i} \frac{b_j^4 + b_k^4 - b_i^4}{4 b_i^2 b_j^2 b_k^2}.    
\end{align}
Replacing Eqs. \eqref{q} and \eqref{Ii'} in Eq. \eqref{field eq R11} we get  the Friedmann equation for our model that is given in terms of the redshift $z$ by
\begin{equation}
\label{H2}
    \frac{H^2}{H_o ^2} = \frac{\displaystyle
        \kappa \left( \Omega_{m o} (1 + z)^3 + \Omega_{\gamma o} (1 + z)^4 \right)  V + \Omega_{K o} (1 + z)^2 - \frac{U}{6 H_o^2}
    }{\displaystyle
        1 - (1 + z) \sum_i I_i - \frac{(1 + z)^2}{6} \left[ \sum_i I_i^2 - \left( \sum_i I_i \right)^2 \right] ,
    }
\end{equation}
where
\begin{align}
\label{U}
    U
&   = - \sum_i \frac{B_{i i}}{2 b_i^2}
    + \sum_{i, j, k} \strconst{^j _k _i} \strconst{^k _j _i} \frac{b_j^4 + b_k^4 - b_i^4}{4 b_i^2 b_j^2 b_k^2},
\\  \Omega_K
&   = \Omega_{K o} (1 + z)^2 \frac{H_o^2}{H^2},
\\  \Omega_{K o}
&   = -\frac{K}{a_o^2 H_o^2}.
\end{align}
Furthermore, from Eqs. $R_{1 k} = 0$ and $R_{i j} = 0$ for $i \neq j$, we have:
\begin{eqnarray}
\label{relation between Ii}
    \sum_l \strconst{^l _k _l} \frac{I_l - I_k}{b_l^2} & = 0,\nonumber\\
\\\label{relation between bi}
    \sum_{k, l} \left[
        \strconst{^k _l _k} \strconst{^l _i _j} \frac{b_l^2 + b_i^2 - b_j^2}{b_k^2}
        +
        \strconst{^k _l _i} \strconst{^l _k _j} \frac{(b_k^2 - b_l^2 + b_i^2) (b_k^2 - b_l^2 - b_j^2)}{b_k^2 b_l^2}
    \right] & = 0,
\end{eqnarray}
Eqs. \eqref{relation between Ii} relates the parameters $I_i$, while Eq. \eqref{relation between bi} relates the parameters $b_i$.\\
In this work we assume a flat-spacetime, then $K = 0$
and since the universe is expanding, that is, $0 < H$.
Also, we suppose that $H$ is a strictly increasing function with respect to redshift, which implies that $0 < H^\prime$.
In consequence, the deceleration parameter has a lower bound
\begin{equation}
\label{lower bound q}
    -1 < q,
\end{equation}
this can be seen from equation (\ref{prime H}). 
Furthermore, we suppose that $b_i$ are the scale factors of the fiber, similar to $a$, then
\begin{equation}
    0 < b_i,
\end{equation}
this implies that $V$ is a positive-valued function.\\
If we replace Eq. \eqref{q} into Eq. \eqref{lower bound q} we obtain
\begin{equation}
    \kappa \left( \frac{3 \Omega_m}{N + 2} + \Omega_\gamma \right) + (1 + z) \sum_i I_i < 3,
\end{equation}
given that $\Omega_m$ and $\Omega_\gamma$ are positive-valued function, then
\begin{equation}
\label{upper bound Ii}
    (1 + z) \sum_i I_i < 3.
\end{equation}

\section{Dark Energy Equation of State, $\omega_e$} 
\label{section:omegae}           
\begin{figure}[hb]
        \centering 
\includegraphics[width=0.5\textwidth]{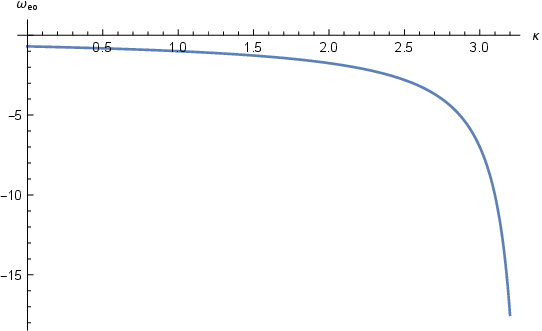}
\caption{Evolution of $\omega_{e 0}$ with respect to $\kappa$, considering
    $q_0 = -0.55$, $\Omega_{m0} = 0.30$ and $\Omega_{\gamma 0} = 10^{-5}$. For these values  and $\kappa=1$, $\omega_{e0}=-1.00019$. }
\label{fig:we_0}
\end{figure}

As in \cite{Hernandez-Marquez_2020}, we consider that the extra terms that appear in the field equations as a consequence of the presence of the fibers can be described as an effective perfect fluid with a barotropic equation of state $p_e=\omega_e\rho_e$. In our model $\rho_e$ is the dark energy that accelerates the 4-dimensional space-time.\\
In order to determine $\omega_e$, we write the analogous equations to Friedmann equations
\begin{align}
\label{G11 eq}
    3 \frac{\dot a^2 + K}{a^2} & = \kappa_h ( \rho + \rho_e ),   \\
\label{G22 eq}
    -2 \frac{\ddot a}{a} - \frac{\dot a ^2 + K}{a^2} & = \kappa_h ( p + p_e) ,   
\end{align}
substituting Eq.\eqref{G11 eq} and \eqref{G22 eq} in the state equation
\begin{equation}
\label{def we}
    \omega_e = \frac{p_e}{\rho_e}
\end{equation}
and again assuming $K = 0$ we get
\begin{equation}
\label{we}
    \omega_e = \frac{1}{3} \frac{2 q - 1 - \kappa \Omega_\gamma}{1 - \kappa ( \Omega_m + \Omega_\gamma )},
\end{equation}
also, we can express $\omega_e$
in terms of $b_i$ and $I_i$.
We divide Eq. \eqref{G11 eq} by critical density \eqref{critical density} to obtain
\begin{equation}
\label{eqdensityparameters}
    \frac{1}{\kappa} = \Omega_m + \Omega_\gamma + \Omega_e.
\end{equation}
Now, we rewrite Eq.\eqref{H2} as Eq. \eqref{eqdensityparameters} and then we compare them to determine
\begin{equation}
    \kappa \Omega_e = \frac{ ( 1 + z ) ^2 }{6} \left[ \sum _i I _i ^2 - \qty( \sum _i I_i ) ^2 \right] + ( 1 + z ) \sum _i I_i  - \frac{U}{6 H^2},
\end{equation}
from
\begin{equation}
    \Omega_e = \frac{\rho_e}{\rho_c}
\end{equation}
we obtain
\begin{equation}
\label{dark density}
    \kappa_h \rho_e = \frac{( 1 + z ) ^2 H^2}{2} \left[ \sum _i I _i ^2 - \qty( \sum _i I_i ) ^2 \right] + 3 ( 1 + z ) H^2 \sum _i I_i  - \frac{U}{2},
\end{equation}
and using Eq. \eqref{q} in Eq. \eqref{G22 eq} we get
\begin{equation}
\label{dark pressure}
    \kappa_h p_e = 3 H^2 \left[ 1 - \kappa \left( \frac{2 \Omega_m}{N + 2} + \Omega_\gamma \right) - \frac{2}{3} (1 + z) \sum_i I_i \right].
\end{equation}
Finally, we substitute Eqs. \eqref{dark density} and \eqref{dark pressure} in Eq. \eqref{def we} to find
\begin{equation}
\label{omegae}
    \omega_e = \dfrac{\displaystyle
        1 - \kappa \left( \dfrac{2 \Omega_m}{N + 2} + \Omega_\gamma \right) - \dfrac{2}{3} (1 + z) \sum_i I_i
    }{\displaystyle
        \dfrac{( 1 + z ) ^2 }{6} \left[ \sum _i I _i ^2 - \qty( \sum _i I_i ) ^2 \right] + ( 1 + z ) \sum _i I_i - \frac{U}{6 H^2}
    }.
\end{equation}
The present value of $\omega_e$, denoted by $\omega_{e o}$, is given by
\begin{equation}
\label{weo}
    \omega_{e 0} = \frac{1}{3} \frac{2 q_0 - 1 - \kappa \Omega_{\gamma 0}}{1 - \kappa ( \Omega_{m 0} + \Omega_{\gamma 0} ) }
\end{equation}
Observe that Eq. \eqref{weo} give us a relation between $\omega_{e o}$ and $\kappa$. 
From equation (\ref{eqdensityparameters}) we have $1- \Omega_{m o} - \Omega_{\gamma o} > \Omega_{e o}$ if $\kappa > 1$ and $1- \Omega_{m o} - \Omega_{\gamma o} < \Omega_{e o}$ if $\kappa < 1$. 
Assuming $\Omega_{m0}=0.3$ and $\Omega_{\gamma0}\approx 10 ^{-5}$, then the current value of $0.7 > \Omega_{e0}$ if $\kappa > 1$ and $0.7 < \Omega_{e o}$ if $\kappa < 1$.
We show this in Figure \ref{fig:omegae} where we plot the evolution of $\Omega_e$ with respect to resdshift for the group $SU(2)$ and the group $SU(3)$ for different values of $\kappa$. Then this equation indicates that $\kappa\approx 1$ to agree with observations, and as a consequence of this  $\kappa_h\approx\kappa_4$.
On the other hand, according to observations $\omega_{e o} < 0$ and $q_o < 0$, then $0 < \Omega_{e o}$, so that
\begin{equation}
\label{upper bound sun constant}
    \kappa < \frac{1}{\Omega_{m o} + \Omega_{\gamma o}}.
\end{equation}
In Figure \ref{fig:we_0} we show the evolution of $\omega_{e o}$ with respect to $\kappa$ considering $q_o = -0.55$, $\Omega_{m o} = 0.3$ and $\Omega_{\gamma o} = 10^{-5}$.
Observe that $\omega_{e o}$ is a decreasing function with respect to $\kappa$ and has a vertical asymptote $\kappa=3.3$. Therefore according to observations $0<\kappa<3.3$. 
Given that $\Omega_{m o}$ and $\Omega_{\gamma o}$ are positive, then $\omega_{e o} < \dfrac{2 q_o - 1}{3}= -0.7$ for $\kappa \in ( 0, 3.3 )$ and $\omega_{e 0} = -1.00019$ at $\kappa = 1$.

\section{One-dimensional Lie Group} 
\label{onedimensional}
In section \ref{section:fieldequations} we found the Friedmann equation for our model given by Eq. (\ref{H2})  and the evolution of the parameters $b_i$ in terms of the redshift $z$. In this section we show that if we consider a one-dimensional Lie-group as a fiber there is no an accelerated expansion of the universe.\\  
In this case, the Eqs. \eqref{q} and \eqref{H2} are reduced to
\begin{align}
\label{q one dim}
    q & = 2 - \kappa \left( \Omega_m + \Omega_\gamma \right) - (1 + z) I,
\\\label{H2 one dim}
    \frac{H^2}{H_0 ^2} & = \frac{
        \kappa \left( \Omega_{m 0} (1 + z)^3 + \Omega_{\gamma 0} (1 + z)^4 \right) \frac{b_0}{b}
    }{1 - (1 + z) I}. 
\end{align}
The numerator of Eq. \eqref{H2 one dim} is a positive-valued function, together with $0 < (H / H_0)^2$ implies that $(1 + z) I < 1$, this means that $H$ is a continuous function, and from Eq. \eqref{H2 one dim} we find
\begin{equation}
\label{I one dim}
    (1 + z) I = 1 - \kappa \left( \Omega_m + \Omega_\gamma \right),
\end{equation}
then putting \eqref{I one dim} into Eq. \eqref{q one dim} we get
\begin{eqnarray}
    q = 1,
\end{eqnarray}
and this implies that if the fiber is a one-dimensional Lie group the four-dimensional spacetime is always decelerating.

\section{One parameter} 
\label{oneparameter}
In this section, we study the behavior of the deceleration parameter considering a Lie group of dimension $N\geq 2$ and we assume that all scale factors are equal, that is, $b_i = b$.
For this case, Eq. \eqref{relation between bi} is reduced to
\begin{equation}
    B_{i j} = \sum_{k,l} \strconst{^k _l _k} \strconst{^l _i _j} \text{ for } i \neq j,
\end{equation}
using Eq. \eqref{Jacobi identity} and then Eq. \eqref{antisymmetry structure constants} we get
\begin{equation}
    B_{i j} = - \sum_{k,l} \left( \strconst{^l _j _k} \strconst{^k _l _i} + \strconst{^l _k _i} \strconst{^k _l _j} \right)
    = - \sum_{k,l} \strconst{^k _l _i} \left(  \strconst{^l _j _k} + \strconst{^l _k _j} \right) = 0  \text{ for } i \neq j,
\end{equation}
this mean that we must choose a basis where the Cartan-Killing metric is diagonal.
Besides, the condition $u_1 = \cdots = u_N$ must be satisfied.
This implies that $B_{i i} = B$.
By the Sylvester’s theorem we have that the Cartan-Killing metric is positive definite or negative definite or zero.
First, we study the case $B \neq 0$.
The field equations are reduced to
\begin{align}
\label{q one parameter}
    q & = 2 - \kappa \qty( \frac{3 \Omega_m}{N + 2} + \Omega_\gamma ) - N ( 1 + z ) I,
\\\label{H2 one parameter}
    \frac{H^2}{H_0^2}
&   = \frac{
    \displaystyle
        \kappa \qty( \Omega_{m o} ( 1 + z )^3 +  \Omega_{\gamma 0} ( 1 + z )^4 ) \left( \frac{b_0}{b} \right)^N
        + \frac{N B}{24 H_o^2 b^2}
    }{
    \displaystyle
        1 - N ( 1 + z ) I + \dfrac{N (N - 1) }{6} ( 1 + z )^2 I^2
    },
\\\label{I' one parameter}
    I^\prime
&   = \qty[
        \kappa \qty( \frac{3 \Omega_m}{N + 2} + \Omega_\gamma ) - 1
    ] \frac{I}{1 + z}
    +  \qty[
        \frac{3 \kappa \Omega_m}{N + 2}
        + \frac{B}{4 H^2 b ^2}
    ] \frac{1}{( 1 + z )^2},
\end{align}
In this model, $U$ is reduced to $U = -\frac{B}{4 b^2}$.
By means of equations we can set the initial conditions at $z = 0$ and solve the system of differential equations numerically to obtain the evolution of $H$, $q$ and the evolution of the scale factor $b$.
\begin{align}
\label{1 parameter I_0}
    I_0 & = \frac{1}{N} \left[ 2 - \kappa \left( \frac{3 \Omega_{m 0}}{N + 2} + \Omega_{\gamma 0} \right)  - q_0 \right]
\\
    b_0
&   = \frac{1}{2 \sqrt{6} H_0} \sqrt{ \frac{-N B}{\kappa ( \Omega _{m 0} + \Omega _{\gamma 0} ) - 1 + N I_0 - \dfrac{N (N - 1)}{6} I_0^2} }.
\end{align}

\begin{figure*}
\includegraphics[width=8cm]{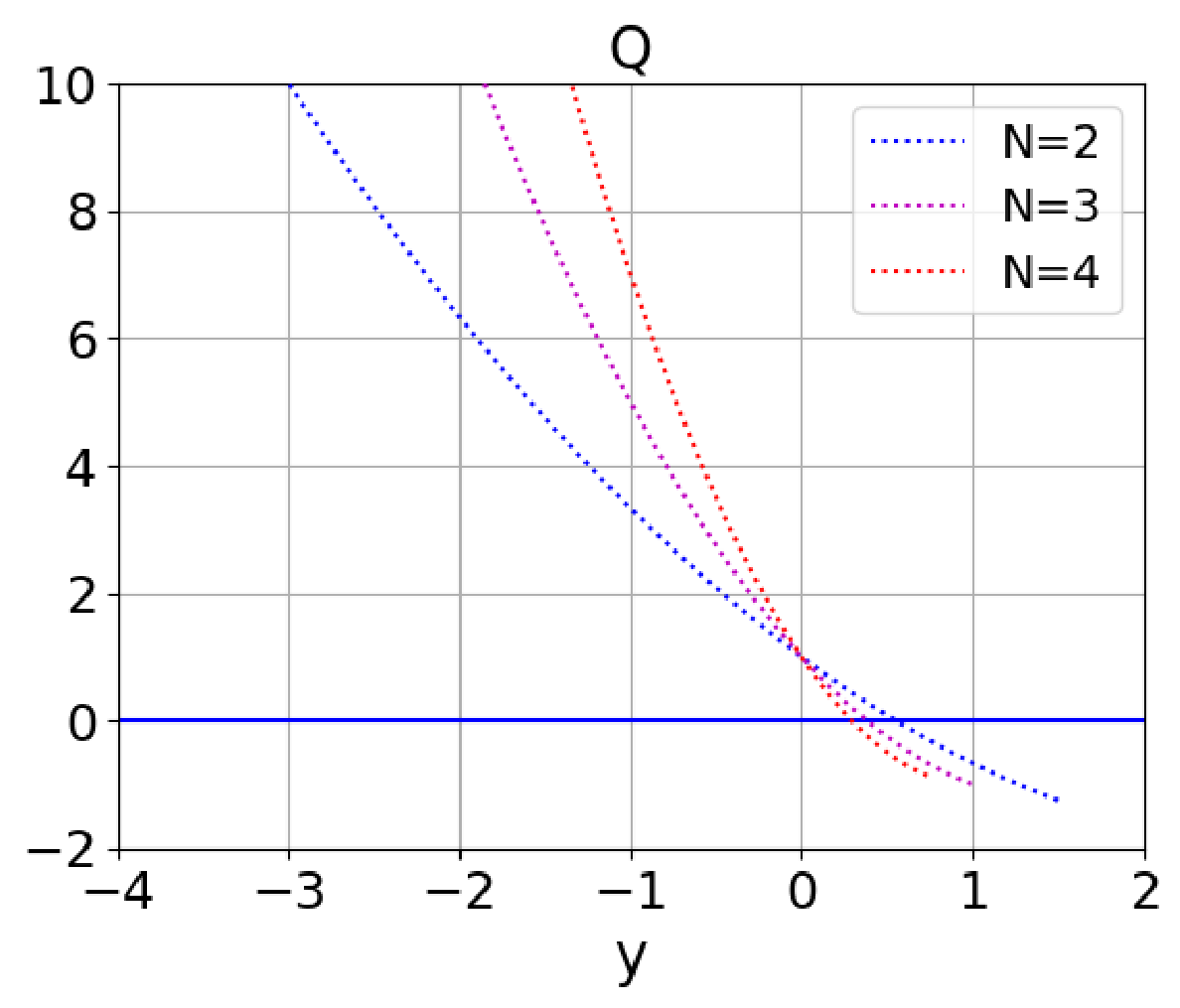}
\caption{$Q$ as function of $y$. The roots of $Q$ for different values of $N$ are shown. The domain of $Q$ is $(-\infty,3/N)$, for $N=2$ the root is at $y=0.55$, for $N=3$ at $y=0.381$, for $N=4$ at $y=0.292$}.
\label{fig: Q vs y}
\end{figure*}

Note that the denominator of Eq. \eqref{H2 one parameter} that we name as $Q$ is a quadratic function which depends on $y\equiv(1 + z) I$. Then $Q=N(N-1)y^2/6-Ny+1$ is a parabola that opens upward, its vertex is $\left( \dfrac{3}{N - 1}, \dfrac{1}{2} \dfrac{N + 2}{N - 1} \right) $ and it crosses the $y$-axis at $y=\dfrac{6/N}{3 \pm \sqrt{3 \dfrac{N + 2}{N}} }$ and according to (\ref{upper bound Ii}) its domain is $y\epsilon(-\infty,3/N)$, then $y = \dfrac{6/N}{3 - \sqrt{3 \dfrac{N + 2}{N}} }$ isn't at the domain of $Q$, $Q$ is an one-to-one function and a strictly decreasing function, see Figure \ref{fig: Q vs y}.
Therefore $H$ can be seen as a function of $y$ and it has an unique jump discontinuity at $y_s = \frac{6}{3N + \sqrt{3N(N + 2)}}$.\\
And since $H^2$ is a positive-valued function, if $y<y_s$ then 
\begin{align}
\label{P positive one parameter}
    -\frac{N B}{24 H^2 b^2} & < \kappa \left( \Omega_m + \Omega_\gamma \right),
\end{align}
and if $y>y_s$
\begin{align}
\label{P negative one parameter}
    -\frac{N B}{24 H^2 b^2} & > \kappa \left( \Omega_m + \Omega_\gamma \right).
\end{align}
It is interesting to note that Eq. \eqref{P negative one parameter} implies that $B < 0$, then the Cartan-Killing metric is negative definite.
Therefore, the model must use a compact semisimple Lie algebra.\\
According to ($\ref{q one parameter}$) we can obtain the image of $q$, the values that can take the deceleration parameter, given by 
\begin{align}
\label{lower bound q one parameter}
    q & > 2 -Ny_s- \kappa \qty( \frac{3 \Omega_m}{N + 2} + \Omega_\gamma ) \hspace{.5cm} for \hspace{.5cm} y<y_s
\\
\label{upper bound q one parameter}
    -1 < q & < 2 -Ny_s - \kappa \qty( \frac{3 \Omega_m}{N + 2} + \Omega_\gamma ),\hspace{.5cm} for \hspace{.5cm} y>y_s
\end{align}
respectively. Then the current value of $q$, has to be in some of these intervals. If we replace $z=0$ in (\ref{lower bound q one parameter}), we obtain $\sqrt{3} - \dfrac{N + 5}{N + 2} - \dfrac{\Omega_{\gamma 0}}{\Omega_{m 0}} < q_0$ and if $N=2$ then $q_0\approx -0.017$, on the other hand if $2 < N$, then $0 < q_0$. Therefore $q_0$ is not in this interval since the observational evidence indicates that the deceleration parameter is negative at the present time and $q_0 \approx -0.55$.\\
Now, if we evaluate Eq. \eqref{upper bound q one parameter} at $z = 0$ obtaining $-1 < q_0 < \dfrac{2 \sqrt{6}}{3 + \sqrt{6}}$ for $1 < N$.
This means that $q_o$ can take the estimated values in the region given by Eq. \eqref{upper bound q one parameter}.
Therefore for any compact semisimple Lie group of dimension $1 < N$ and considering equal parameters in the metric, the current value of $q_0$ can be negative, then it is possible to obtain an accelerated expansion at present time. Once we know that the current value of the deceleration parameter  $q_0$ is in the interval given by (\ref{upper bound q one parameter}), we want to know if it is an increasing function with respect to redshift.
Then if we compute its first derivative with respect to redshift, we obtain: 
\begin{equation}
\label{derivative q}
    q^\prime = -\kappa \qty( \frac{3 \Omega_m^\prime}{N + 2} + \Omega_\gamma^\prime ) - N y^\prime,
\end{equation}
from the Eq. \eqref{I' one parameter} we have
\begin{equation}
\label{(1 + z) I' + I}
    y^\prime
    = \kappa \qty( \frac{3 \Omega_m}{N + 2} + \Omega_\gamma ) I
    +  \qty[
        \frac{3 \kappa \Omega_m}{N + 2}
        + \frac{B}{4 H^2 b ^2}
    ] \frac{1}{1 + z},
\end{equation}
while the derivatives of density parameters with respect to redshift are
\begin{align}
\label{derivative density parameter matter}
    \Omega_m^\prime & = \frac{\Omega_m}{1 + z} \left[ -1 + \kappa \left( \frac{3 \Omega_m}{N + 2} + \Omega_\gamma \right) - q \right],
\\\label{derivative density parameter radiation}
    \Omega_\gamma^\prime & = \frac{\Omega_\gamma}{1 + z} \left[ \kappa \left( \frac{3 \Omega_m}{N + 2} + \Omega_\gamma \right) - q \right],
\\\label{derivative density parameter radiation matter}
    & = \frac{\Omega_\gamma}{\Omega_m} \Omega_m^\prime + \frac{\Omega_\gamma}{1 + z}.
\end{align}
Observe that the derivatives of $I$ and the density parameters are not defined at $y_s$.
Replacing Eqs. \eqref{(1 + z) I' + I} and \eqref{derivative density parameter radiation matter} in Eq. \eqref{derivative q} we obtain
\begin{equation}
    q^\prime = -3\frac{N + 1 - 2 q}{N + 2} \frac{\kappa \Omega_m}{1 + z} - 2 (1 - q) \frac{\kappa \Omega_\gamma}{1 + z} - \frac{N B}{4 (1 + z) H^2 b^2}.
\end{equation}
Given that $3\dfrac{N + 1 - 2 q}{N + 2}<6$ and $2 (1 - q)<6$ then
\begin{equation}
    (1 + z) q^\prime > -6 \kappa \left( \Omega_m + \Omega_\gamma \right) - \frac{N B}{4 H^2 b^2},
\end{equation}
and if $y>y_s$ then according to (\ref{P negative one parameter}) $q'>0$. But if $y<y_s$ then (\ref{P positive one parameter}) is satisfied and $q^\prime$ is positive as long as
\begin{equation}
\label{P and q' positive}
    3\frac{N + 1 - 2 q}{N + 2} \Omega_m + 2 (1 - q)  \Omega_\gamma < - \frac{N B}{4 \kappa H^2 b^2}.
\end{equation}
This means that if we consider any compact semisimple Lie group of dimension $1 < N$, then $q$ is a strictly increasing function with respect to $z$.
Physically this means that for any compact semisimple Lie group there is a current accelerated expansion $q_0<0$ but as $z$ increases this accelerated expansion slows down and at some redshift $z$ has to change to $q>0$. Therefore there is a redshift when the universe starts to accelerate.\\

In the case $B = 0$ Eq. \eqref{H2 one parameter} is reduced to
\begin{equation}
\label{H2 one parameter B zero}
    \frac{H^2}{H_0^2} = \frac{
    \displaystyle
        \kappa \qty( \Omega_{m o} ( 1 + z )^3 +  \Omega_{\gamma 0} ( 1 + z )^4 ) \left( \frac{b_0}{b} \right)^N
    }{
    \displaystyle
        1 - N ( 1 + z ) I + \dfrac{N (N - 1) }{6} ( 1 + z )^2 I^2
    }
\end{equation}
Given that $( H / H_o )^2$ and its numerator in Eq. \eqref{H2 one parameter B zero} are positive functions, then its denominator is also a positive function, so that $q$ satisfies Eq. \eqref{lower bound q one parameter}.
We found that $q_o$ do not agree with the estimated values.
\section{$SU(\mathscr{N})$} 
\label{section:SUN}
\begin{figure}
\includegraphics[width=18cm]{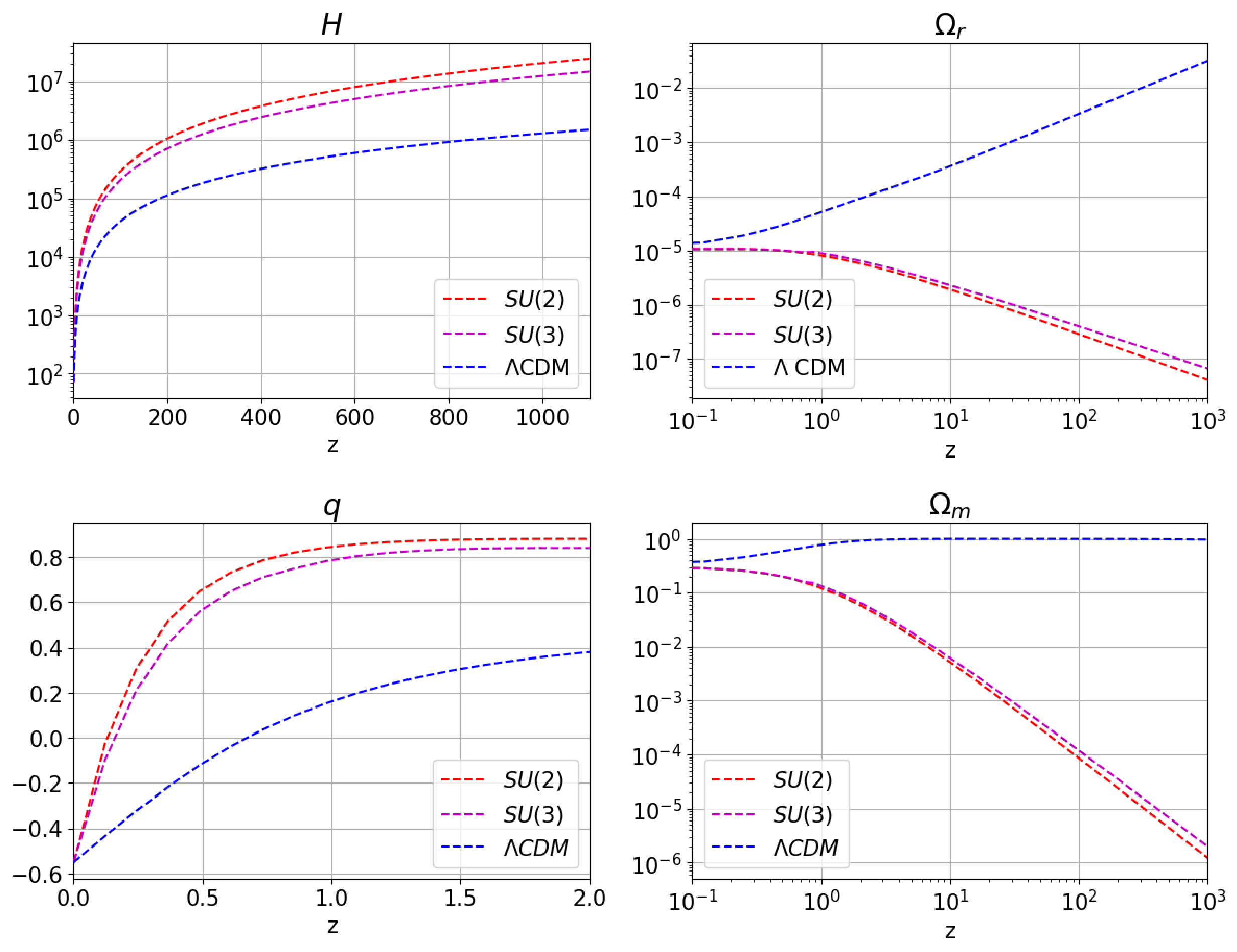}
\caption{Comparison of the Hubble parameter $H$, deceleration parameter $q$ and density parameters $\Omega_m$ and $\Omega_{\gamma}$ with the $\Lambda$CDM model, considering $SU(2)$ and $SU(3)$ as fibers. From this figure we can see that $q$ is an increasing function with respect to $z$. We can see that if the fiber is the group $SU(2)$ the universe starts to accelerate at $z\approx0.1$ and for the group $SU(3)$ at $z\approx 0.157$.}
\label{parametersSUN}
\end{figure}

In this section we study the case where the fiber is the group $SU(\mathscr{N})$ with $1 < \mathscr{N}$.
The dimension of this group is $N = \mathscr{N}^2 - 1$.
The structure constants of $SU(\mathscr{N})$ are antisymmetric in all indices and its corresponding Cartan-Killing metric is $B_{i j} = - \mathscr{N} \delta_{i j}$.
Furthermore, we assume that all scale factors are the same, that is, $b_i = b$. The field equations are changed to
\begin{align}
\label{q sun}
    q & = 2 - \kappa \qty( \frac{3 \Omega_m}{\mathscr{N}^2 + 1} + \Omega_\gamma ) - ( \mathscr{N}^2 - 1 ) ( 1 + z ) I,
\\\label{H2 sun}
    \frac{H^2}{H_o^2}
&   = \frac{
    \displaystyle
        \kappa \qty( \Omega_{m o} ( 1 + z )^3 +  \Omega_{\gamma o} ( 1 + z )^4 ) \left( \frac{b_o}{b} \right)^{\mathscr{N}^2 - 1}
        - \frac{\mathscr{N} (\mathscr{N}^2 - 1)}{24 H_o^2 b^2}
    }{
    \displaystyle
        1 - ( \mathscr{N}^2 - 1 ) ( 1 + z ) I + \dfrac{ ( \mathscr{N}^2 - 1 ) ( \mathscr{N}^2 - 2 ) }{6} ( 1 + z )^2 I^2
    },
\\\label{I' sun}
    I^\prime
&   = \qty[
        \kappa \qty( \frac{3 \Omega_m}{\mathscr{N}^2 + 1} + \Omega_\gamma ) - 1
    ] \frac{I}{1 + z}
    +  \qty[
        \frac{3 \kappa \Omega_m}{\mathscr{N}^2 + 1}
        - \frac{\mathscr{N}}{4 H^2 b ^2}
    ] \frac{1}{( 1 + z )^2},
\end{align}
For this case if we set $z=0$ in the above equations we can find $I_0$ and $b_0$ in terms of $\Omega_{m0},\Omega_{\gamma 0}$, $q_0$ and $H_0$.\\
\begin{align}
\label{Io sun}
    I_0 & = \frac{1}{\mathscr{N}^2 -1} \left( 2 - \kappa \left( \frac{3 \Omega_{m 0}}{\mathscr{N}^2 + 1} + \Omega_{\gamma 0} \right)  - q_0 \right)
\\
\label{b0sun}
    b_0
&   = \frac{1}{2 \sqrt{6} H_0} \sqrt{ \frac{\mathscr{N} ( \mathscr{N}^2 - 1 )}{\kappa ( \Omega _{m 0} + \Omega _{\gamma 0} ) - 1 + ( \mathscr{N}^2 - 1 ) I_0 - \dfrac{ ( \mathscr{N}^2 - 1 ) ( \mathscr{N}^2 - 2 ) }{6} I_0^2} }.
\end{align}
 In order to obtain the evolution of the Hubble parameter $H$, $q$ and the density parameters, we solve numerically (\ref{q sun}), (\ref{H2 sun}) and (\ref{I' sun}) using the above initial conditions (\ref{Io sun}), (\ref{b0sun}) at $z=0$ for $I_0$ and $b_0$ .
We show these numerical solutions in  Figure \ref{parametersSUN} for the $SU(2)$ and the $SU(3)$ group.

\begin{figure}
    \centering
    \includegraphics[width=8cm]{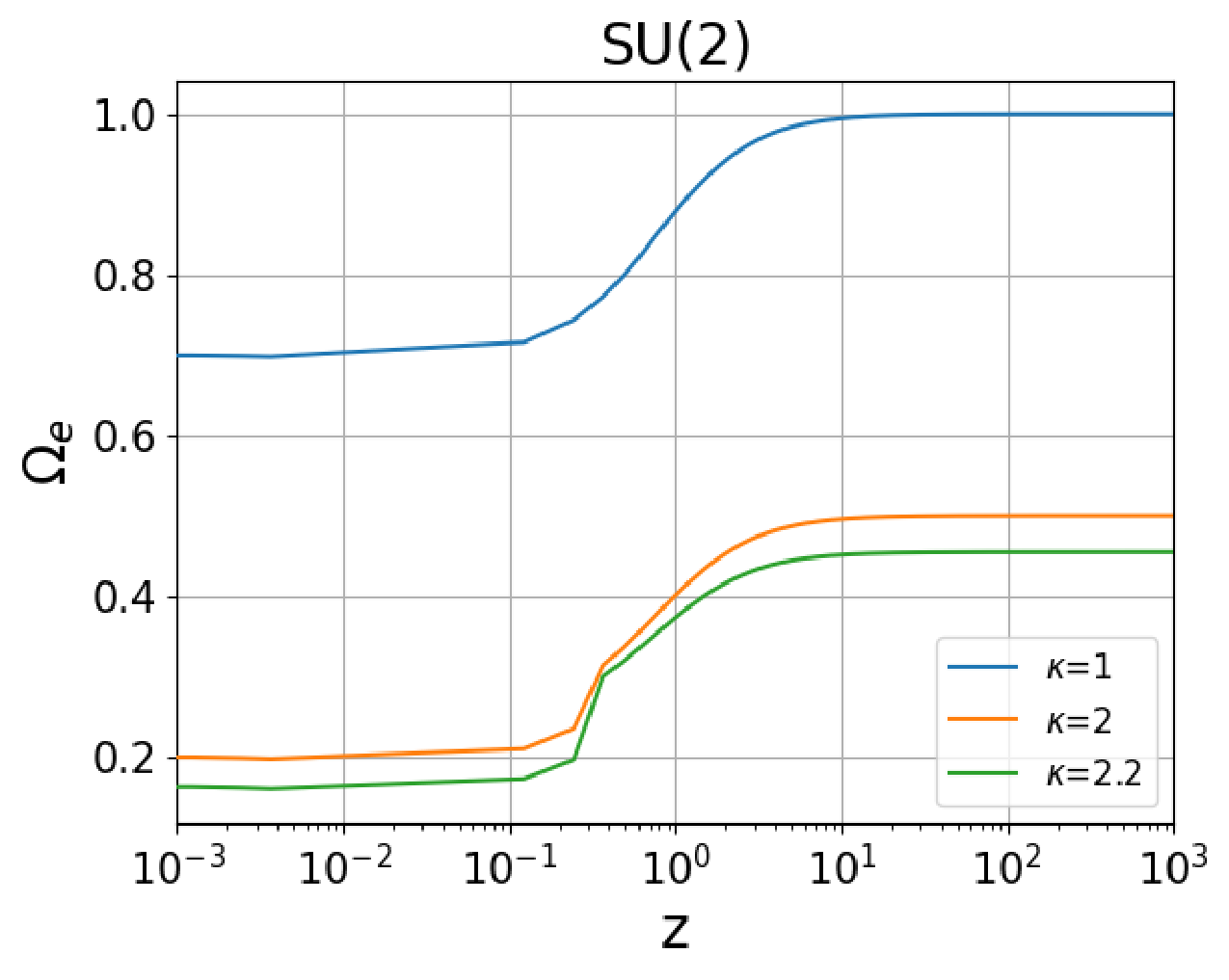}\includegraphics[width=8cm]{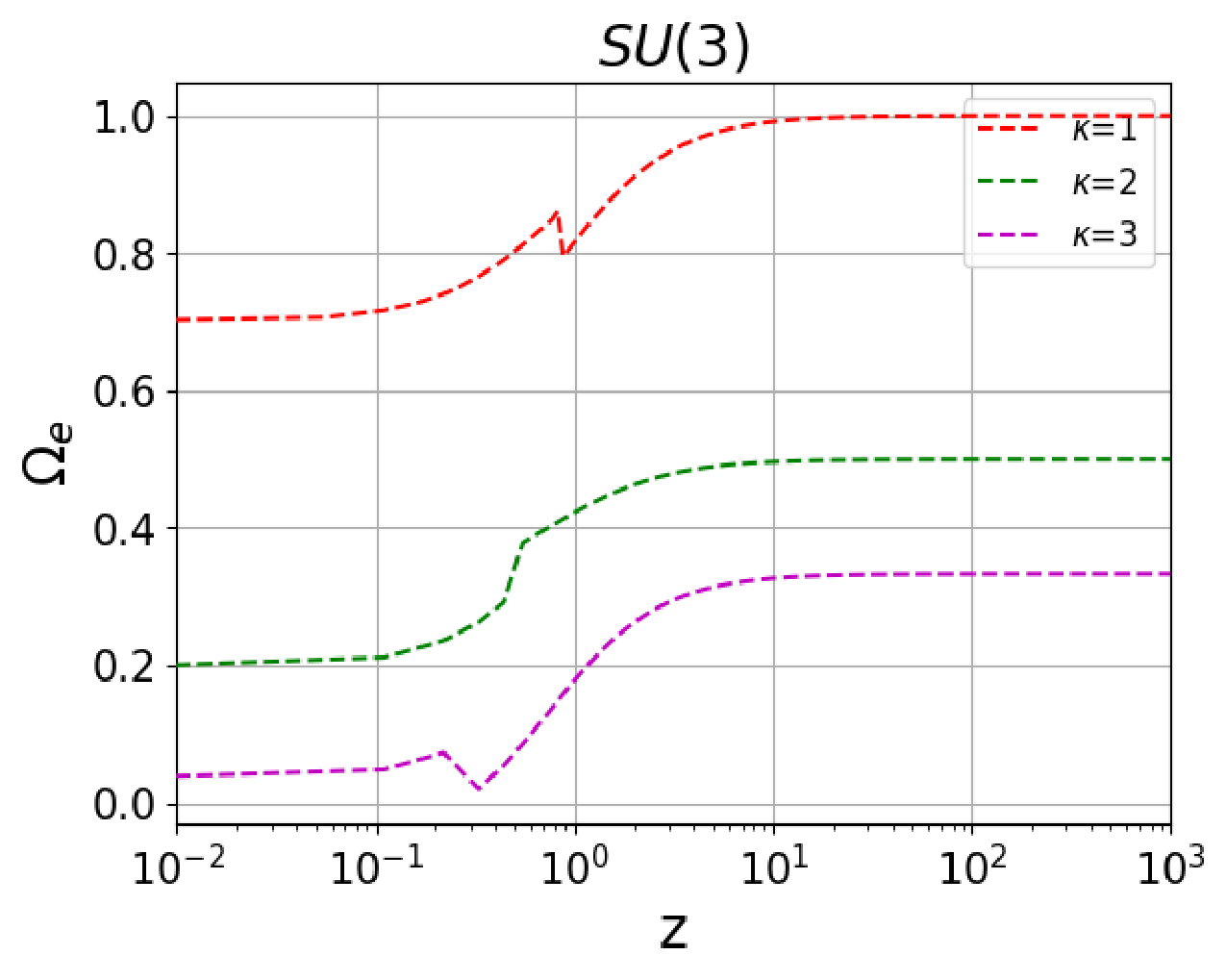}
    \caption{Evolution of $\Omega_e$ for the group $SU(2)$ and $SU(3)$, respectively. In both cases the parameter of density of dark energy increases with respect to redshift. }
    \label{fig:omegae}
\end{figure}

\begin{figure}
    \centering
    \includegraphics[width=8cm]{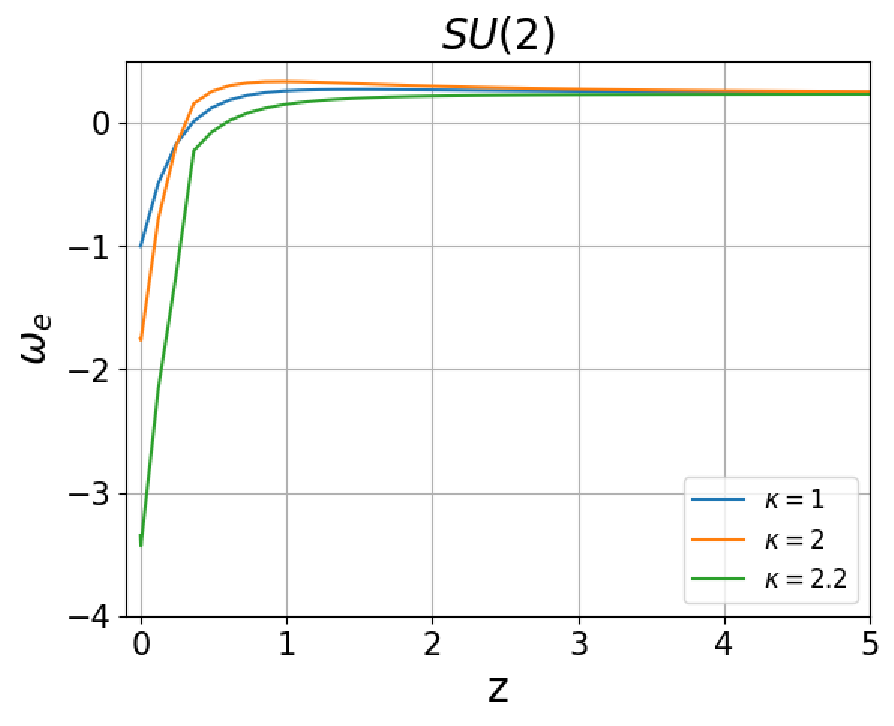}\includegraphics[width=8cm]{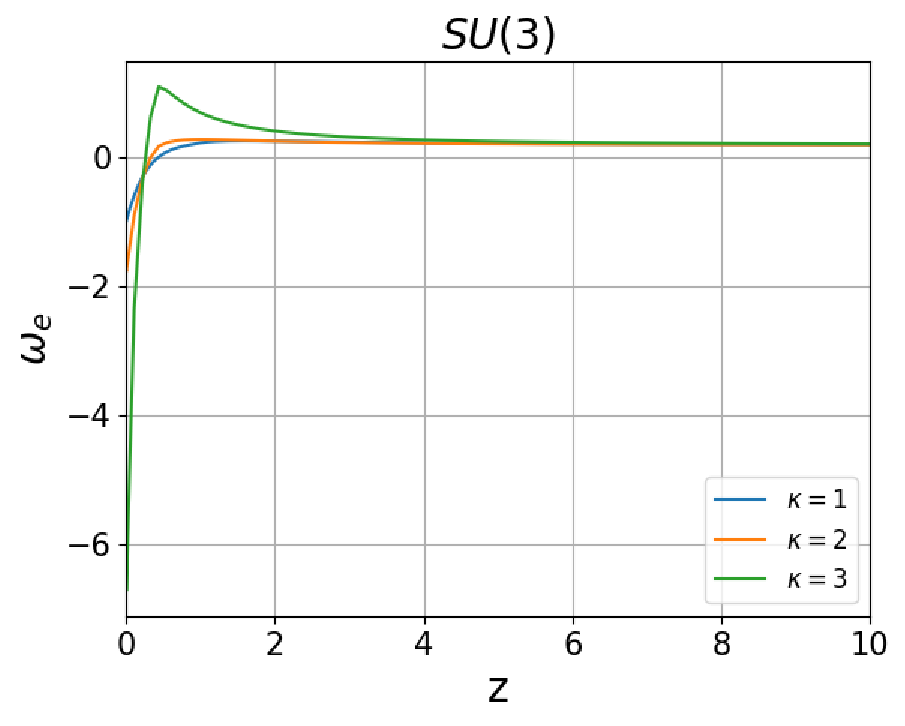}
    \caption{Evolution of $\omega_e$ with respect to redshift for different valuesFigure of $\kappa$. We can see that if the fiber for the principal bundle is the group $SU(2)$ and $\kappa=1$, then the current value of $\omega_e=-1.00019$ and is the same value for $SU(3)$. }
    \label{figwe}
\end{figure}

\section{Lie group with different scale factors}  
\label{section9:differentscale}                   
In this case we consider a Lie group with $1 < N$, where the scale factors that appears in the metric are different. Then $H^2$ is given by Eq. (\ref{H2}) and we first consider the case $U \neq 0$. As before the denominator is a function of $y\equiv(1+z)\sum_iI_i(z)<3$ , then $H$ is discontinuous at $y_s$, where $y_s$ is given by
\begin{equation}
   y_s\equiv (1 + z_s) \sum_i I_i(z_s) = 3 - \sqrt{(1 + z_s)^2 \sum_i I_i(z_s)^2 + 3},
\end{equation}
where $z_s$ is the value of the redshift $z$ that satisfies this equality.\\
Let $P$ be the numerator of $(H/H_o)^2$ in Eq. (\ref{H2}). If $y>y_s$, this implies $P<0$ and therefore
\begin{equation}
\label{P less 0}
    \kappa ( \Omega_m + \Omega_\gamma )  < \frac{U}{6 H^2},
\end{equation}
and if $y<y_s$ , then $P>0$  and  
\begin{equation}
        \frac{U}{6 H^2}  < \kappa ( \Omega_m + \Omega_\gamma ).
\end{equation}
Note that Eq. \eqref{P less 0} implies that
\begin{equation}
    0 < U \text{ when }  y>y_s,
\end{equation}
therefore, the structure constants together with the scale factors of the fiber must satisfy this condition. If we observe Eq. (\ref{U}), we note that we must use a non-Abelian Lie group for this model.\\
And according to Eq. \eqref{q} we get
\begin{eqnarray}
\label{qdifferent1}
   -1<q< 2 - \kappa \left( \frac{3 \Omega_m}{N + 2} + \Omega_\gamma \right)-y_s  \hspace{.5cm} &for \hspace{.5cm} y>y_s,\\
\label{qdifferent2}
2- \kappa \left( \frac{3 \Omega_m}{N + 2} + \Omega_\gamma \right)-y_s< q \hspace{.5cm} &for\hspace{.5cm} y<y_s.
\end{eqnarray}
Similar to the previous section, the present value of the deceleration parameter $q_0$ is negative at the interval given by \eqref{qdifferent1}.
And if we compute $q'$, we obtain
\begin{equation}
\label{q'}
    q^\prime = - \kappa \left( \frac{3 \Omega_m^\prime}{N + 2} + \Omega_\gamma^\prime \right) - \sum_i \left[ (1 + z) I_i^\prime + I_i \right],
\end{equation}
from Eq. \eqref{Ii'} we find
\begin{equation}
\label{derivative (1+z)I}
    \sum_i \left[ (1 + z) I_i^\prime + I_i \right] = \kappa \left( \frac{3 \Omega_m}{N + 2} + \Omega_\gamma \right) \sum_i I_i + \left( \frac{3 \kappa N\Omega_m}{N + 2} - \frac{U}{H^2} \right) \frac{1}{1 + z},
\end{equation}
and differentiating Eq. \eqref{def density parameter} with respect to $z$, we get
\begin{equation}
\label{density parameter derivative}
    \Omega^\prime = \frac{\Omega}{1 + z} \left[ 3 \omega - 1 + \kappa \left( \frac{3 \Omega_m}{N + 2} + \Omega_\gamma \right) - q \right].
\end{equation}
Substituting Eq. \eqref{derivative (1+z)I} into Eq. \eqref{q'} and using Eq. \eqref{density parameter derivative} we find
\begin{equation}
    q^\prime = -3 \frac{N + 1 - 2 q}{N + 2} \frac{\kappa \Omega_m}{1 + z} -2 (1 - q) \frac{\kappa \Omega_\gamma}{1 + z} + \frac{U}{(1 + z) H^2},
\end{equation}
For $y>y_s$ we have
\begin{equation}
    2 \kappa \left( \frac{3 \Omega_m}{N + 2} + \Omega_\gamma \right) \frac{2 + q}{1 + z} < q^\prime,
\end{equation}
then, $0 < q^\prime$, therefore, from \eqref{qdifferent1} we have that the current value of $q$ can be negative for $y>y_s$ and from the last expression that is an increasing function with respect to redshift for $y>y_s$, then for $y>y_s$ the deceleration parameter grows from $q_0$ with respect to redshift. 
And when $y<y_s$, $q$ is an strictly increasing function as long as
\begin{equation}
    \frac{N + 1 - 2 q}{N + 2} \frac{\kappa \Omega_m}{2} + (1 - q) \frac{\kappa \Omega_\gamma}{3}  < \frac{U}{6 H^2},
\end{equation}
is verified.\\

Now, we assume that $U = 0$, then
\begin{equation}
    0 < 1 - (1 + z) \sum_i I_i - \frac{(1 + z)^2}{6} \left[ \sum_i I_i^2 - \left( \sum_i I_i \right)^2 \right],
\end{equation}
so that $y < 3 - \sqrt{3}$.
Using Eq. \eqref{q} we find that
\begin{equation}
\label{lower bound q U zero}
    \sqrt{3} - 1 - \kappa \left( \frac{3 \Omega_m}{N + 2} + \Omega_\gamma \right) < q
\end{equation}
Evaluating Eq. \eqref{lower bound q U zero} at $z = 0$ we obtain $\sqrt{3} - \dfrac{N + 5}{N + 2} - \dfrac{\Omega_{mo}}{\Omega_{\gamma o}} < q_o$.
Before, we found that these values of $q_o$ do not agree with the observations.
\section{Evolution of density parameters.}  
\label{section:evolutiondensity}
According to (\ref{derivative density parameter matter}) and (\ref{derivative density parameter radiation})  the evolution of the matter and radiation density parameters is determined by the function
\begin{equation}
\label{F}
    F = -1 + \kappa \left( \frac{3 \Omega_m}{N + 2} + \Omega_\gamma \right) - q,
\end{equation}
if $F>0$ then $\Omega_m$ is an increasing function with respect to $z$ but if $F<0$ then $\Omega_m$ is a decreasing function with respect to $z$.  
If we interpret to $\rho_e$ as a real physical density, we have $0 < \Omega_e$, then $\kappa \left( \dfrac{3 \Omega_m}{N + 2} + \Omega_\gamma \right) < 1$, so that $F < -q$.
For $0 < q$ we have that $F < 0$, therefore when the universe is slowing down the matter density parameter decreases with respect to z.\\
On the other hand, $F$ is continuous when $q_0 \leq q \leq 0$ for both models (one parameter and several parameters).
Computing the derivative of $F$ we find
\begin{equation}
    (1 + z) F^\prime = \frac{3 \kappa \Omega_m}{N + 2} \left[ N - 3q + \kappa \left( \frac{3 \Omega_m}{N + 2} + \Omega_\gamma \right) \right]
    + \kappa \Omega_\gamma \left[ 2 - 3q + \kappa \left( \frac{3 \Omega_m}{N + 2} + \Omega_\gamma \right) \right] - \frac{U}{H^2},
\end{equation}
If $q_0 \leq q \leq 0$ then $q$ satisfies $\kappa ( \Omega_m + \Omega_\gamma )  < \frac{U}{6 H^2}$ for both models, so that
\begin{equation}
\label{F' Region I}
    (1 + z) F^\prime < \kappa \left( \frac{3 \Omega_m}{N + 2} + \Omega_\gamma \right) \left[
        \kappa \left( \frac{3 \Omega_m}{N + 2} + \Omega_\gamma \right) - 4 - 3 q
    \right]
    < - 3 \kappa \left( \frac{3 \Omega_m}{N + 2} + \Omega_\gamma \right) ( 1 + q ) < 0
\end{equation}
Therefore, $F^\prime < 0$.
This implies that $F$ is a strictly decreasing function when the universe is accelerating.
If we compute the current value of $F_0$, then $F_0=-1-q_0+3\Omega_{m0}/(N+2)+\Omega_{\gamma0}$ and with the current values of $\Omega_{m0}=0.3$, $q_0=-.55$, $\Omega_{\gamma0}=10^{-5}$, we obtain that for any compact semisimple Lie group for a parameter o non-Abelian Lie group for several parameters of dimension $1 < N$, $F_0$ is negative. 
This means $\Omega_m$ decreases from $\Omega_{m 0}$ to approach to 0.
According to (\ref{derivative density parameter radiation}) we can rewrite $\Omega_{\gamma}'$ as $\Omega_\gamma^\prime = \dfrac{\Omega_\gamma}{1 + z} (1 + F)$.
$\Omega_\gamma^\prime$ is positive when $-1 < F$, it is zero in $F = -1$ and it is negative when $F < -1$.
Note that the present value of $F$, satisfies $-1 < -(1 + q_0) < F_0$ independently of the values of $N$ and $\kappa$.
Therefore, the radiation density parameter grows from its current value $\Omega_{\gamma 0}$ to its maximum value and then it decreases.
\section{Conclusions}
\label{conclusions}
In this work we show that if we consider the topology of our universe as a principal fiber bundle whose base-space is a homogeneous and isotropic 4-dimensional space-time described by the F-R-W metric and the fiber is an $N-$dimensional Lie group; and if we assume that the $N+4$-dimensional Einstein field equations are valid then the Hubble parameter, the deceleration parameter and the density parameters are affected by the presence of the fibers.\\
Furthermore, we found  for $N>1$, that if we consider equal scale-factors of the fiber then the base space is currently accelerating, $q_o<0$, if the fiber is a compact semi-simple Lie group but when we consider different scale factors of the fiber then the base space is currently accelerating if the fiber is a non-Abelian Lie group as long as its structure constants satisfy $U>0$. Also we show that $q$ is an increasing function with respect to $z$ meaning that in our model, the current accelerated expansion at $z=0$, $q_0<0$ slows down as $z$ increases and finishes at some redshift $z$, where the  expansion of the universe starts to slows down.\\
In section \ref{section:omegae} we define the equation of state of what we call "dark energy" in our model and we show that the current value of $\omega_{e o}$ is negative for $N>1$ compact semi-simple Lie groups whose scale-factors are equal and for $N>1$ non-Abelian Lie groups whose scale-factors are different. We also find that the ratio between the $4$-dimensional gravitational constant $\kappa_4$ and the $N+4$-dimensional gravitational constant $\kappa_h$ and  considering the current observed values of matter and radiation $\Omega_{m o}=0.3$ and $\Omega_{\gamma0}\approx 10^{-5}$ it is constrained to some values, that is to say, $0<\kappa<3.3$ and when $\kappa=1$ the current value of $\omega_{e0}=-1.00019$, therefore $\kappa\approx 1$ to agree with observations and $\kappa_4\approx\kappa_h$.\\
In section \ref{section:SUN} we find the initial conditions at $z=0$ for the group $SU(\mathcal{N})$. These initial conditions depend only on the dimension of the group and on the current values of deceleration parameter and density parameters, with these initial conditions we solve the set of equations (\ref{q sun}),(\ref{H2 sun}), (\ref{I' sun}), for the group $SU(2)$ and the group $SU(3)$. For both groups we show the evolution of the Hubble parameter, deceleration parameter, density parameters in Figure \ref{parametersSUN} and dark energy equation of state $\omega_e$ in Figure \ref{figwe}.\\
However although the $N>1$ Lie groups described above allows an accelerated current expansion of the base-space the evolution of the density parameters don't evolve according to the $\Lambda$CDM model. This is because according to section \ref{section:evolutiondensity}, the density parameter of matter decreases from its current value $\Omega_{m0}$ with respect to redshift and tends to zero while the density parameter of radiation grows from $\Omega_{\gamma 0}$ to its maximum value and then it decreases with respect to redshift. This can be seen in Figure \ref{parametersSUN} where we show the evolution of density parameters for the group $SU(2)$ and the group $SU(3)$. On the other hand the density parameter of dark energy grows from its current value $\Omega_{e0}$ at $z=0$ to $1$ for $z\approx 1000$, in the case for $\kappa=1$. We can see this in Figure \ref{fig:omegae} when we also include the evolution of $\Omega_{e}$ for different values of $\kappa$.\\
Finally, the main conclusion is that the accelerated expansion of the universe observed today is probably due to the topology of the universe, rather than to the presence of an exotic type of matter or a modification of Einstein's equations. Unfortunately, in the model proposed here the densities obtained do not follow the observed evolution of the universe, but this could be due to the choice of the fiber group or to another topological configuration that should be explored in the future.
\section*{Acknowledgments}
We acknowledge financial
support from CONACYHT postdoctoral fellowships. 
\printbibliography
\end{document}